\documentclass[preprint,prd,aps,eqsecnum,nofootinbib]{revtex4}
\usepackage{amsmath}

\begin{document}

\def\cstok#1{\leavevmode\thinspace\hbox{\vrule\vtop{\vbox{\hrule\kern1pt
\hbox{\vphantom{\tt/}\thinspace{\tt#1}\thinspace}}
\kern1pt\hrule}\vrule}\thinspace}

\begin{center}
\bibliographystyle{article}
{\Large \textsc{Casimir apparatuses in a weak gravitational field}}
\end{center}

\author{Giuseppe Bimonte$^{3,1}$ \thanks{Electronic address:
giuseppe.bimonte@na.infn.it},
Enrico Calloni$^{3,1}$ \thanks{Electronic address:
enrico.calloni@na.infn.it},
Giampiero Esposito$^{1}$ \thanks{Electronic address:
giampiero.esposito@na.infn.it}, 
George M. Napolitano$^{2}$ \thanks{Electronic address:
gmn@math.ku.dk},
Luigi Rosa$^{3,1}$ \thanks{Electronic address: 
luigi.rosa@na.infn.it}} 

\affiliation{
${\ }^{1}$Istituto Nazionale di Fisica Nucleare, Sezione di Napoli,\\
Complesso Universitario di Monte S. Angelo, Via Cintia, Edificio 6, 80126
Napoli, Italy\\
${\ }^{2}$Department of Mathematics, Copenhagen University, 
Universitetsparken 5, DK-2100 Copenhagen, Denmark\\
${\ }^{3}$Dipartimento di Scienze Fisiche, Complesso Universitario di Monte
S. Angelo,\\
Via Cintia, Edificio 6, 80126 Napoli, Italy}

\vspace{0.4cm}
\date{\today}

\begin{abstract}
We review and assess a part of the 
recent work on Casimir apparatuses in the weak
gravitational field of the Earth. For a free, real massless scalar field
subject to Dirichlet or Neumann boundary conditions on the parallel plates,
the resulting regularized and renormalized energy-momentum tensor is
covariantly conserved, while the trace anomaly vanishes if the massless
field is conformally coupled to gravity. Conformal coupling also ensures a
finite Casimir energy and finite values of the pressure upon parallel
plates. These results have been extended to an electromagnetic field subject
to perfect conductor (hence idealized) boundary conditions
on parallel plates, by various authors. The regularized and renormalized
energy-momentum tensor has been evaluated up to second order in the gravity
acceleration. In both the scalar and the electromagnetic case, studied to
first order in the gravity acceleration, the theory predicts a tiny force in
the upwards direction acting on the apparatus.
This effect is conceptually very interesting, since it means that Casimir
energy is indeed expected to gravitate, although the magnitude of the
expected force makes it necessary to overcome very severe signal-modulation
problems.
\end{abstract}

\maketitle
\bigskip
\vspace{2cm}

\section{Introduction}

Ever since Casimir discovered that suitable differences of zero-point
energies of the quantized electromagnetic field can be made finite and
provide measurable effects \cite{KNAWA-51-793}, 
several efforts have been produced to understand
the physical implications and applications of this 
property \cite{PRPLC-353-1}--\cite{IJSQE-13-400}. In particular,
we are here going to review the recent theoretical discovery that Casimir
energy gravitates \cite{PHRVA-D76-025004}-\cite{PHRVA-D76-025008}. 
In Ref. \cite{PHRVA-D74-085011}, 
this was proved as part of an investigation that led, for the first time, 
to the evaluation of the energy-momentum tensor of a Casimir
apparatus in a weak gravitational field (cf. the work in Ref. 
\cite{PHRVA-D69-085005}). In that piece of work, Maxwell
theory was quantized via functional integral, with perfect conductor 
boundary conditions on parallel plates at distance $a$ from each other.
On using Fermi--Walker coordinates, where the $(x_{1},x_{2})$ coordinates
span the plates, while the $z=x_{3}$ axis coincides with the vertical 
upward direction (so that the plates have equations $z=0$ and $z=a$,
respectively), and working to first order in the constant gravity
acceleration $g$, the spacetime metric reads as
\cite{PHRVA-D74-085011}
\begin{equation}
ds^{2}=-c^{2} \left(1+{\varepsilon}{z\over a}\right)dt^{2}
+dx_{1}^{2}+dx_{2}^{2}+dz^{2}+{\rm O}(|x|^{2}),
\label{(1.1)}
\end{equation}
where $\varepsilon \equiv {2ga \over c^{2}}$.

Our paper provides a review of some key findings by the authors and
by other research groups interested in the same topics. For this purpose,
Sec. II studies the Feynman Green function for the scalar wave operator
to zeroth and first order in $\varepsilon$, Sec. III obtains the 
resulting regularized and renormalized energy-momentum tensor while 
Sec. IV evaluates Casimir energy and pressure upon the plates. All of this
with Dirichlet conditions on the plates for the Green function. The case 
of Neumann boundary conditions is considered in Sec. V, while the 
electromagnetic analysis is summarized in Sec. VI. Concluding remarks
are presented in Sec. VII.

\section{Feynman Green function to zeroth and first order}
  
To first order in the $\varepsilon$ parameter of Sec. I, 
the only nonvanishing Christoffel symbols associated with the metric
(1.1) are 
\begin{equation}
\Gamma_{\; 30}^{0}=\Gamma_{\; 03}^{0}
={\varepsilon \over 2 (a+\varepsilon z)}
\sim {\varepsilon \over 2a}+{\rm O}(\varepsilon^{2}), \;
\Gamma_{\; 00}^{3} \sim {\varepsilon \over 2a}
+{\rm O}(\varepsilon^{2}).
\label{(2.1)}
\end{equation}
We now compute the wave operator $\cstok{\ }$,
the Feynman Green function of the hyperbolic operator
$(\cstok{\ }- \xi R)$, and eventually the Hadamard function and the
regularized energy-momentum tensor.

Indeed, a Green function of the scalar wave operator 
obeys the differential equation
\begin{equation}
(\cstok{\ }-\xi R)G(x,x')=-{\delta(x,x')\over \sqrt{-g}}.
\label{(2.2)}
\end{equation}
The Feynman Green function $G_{F}$ is the unique symmetric complex-valued
Green function which obeys the relation \cite{DeWi65}
$$
\delta G = G \; \delta F \; G,
$$
where $F$ is the invertible operator obtained from variation of the
action functional with respect to the field. This definition is well
suited for the purpose of defining the Feynman Green function even when
asymptotic flatness does not necessarily hold \cite{DeWi65}.

In our first-order expansion in the $\varepsilon$ parameter, the
scalar curvature gives vanishing contribution to Eq. (2.2), which
therefore takes the form (hereafter $\cstok{\ }^{0} \equiv
\eta^{\mu \nu}\partial_{\mu} \partial_{\nu}$)
\begin{equation}
\left(\cstok{\ }^{0}+{\varepsilon z \over (a+ \varepsilon z)}
{\partial^{2}\over \partial t^{2}}
+\Gamma_{\; 00}^{3}{a\over (a+\varepsilon z)}
{\partial \over \partial z} \right)G(x,x')
=-{\delta(x,x')\over \sqrt{-g}}.
\label{(2.3)}
\end{equation}
We now follow our work in Ref. \cite{PHRVA-D74-085011} and assume
that the Feynman Green function admits the asymptotic expansion
\begin{equation}
G_{F}(x,x') \sim G^{(0)}(x,x')+\varepsilon G^{(1)}(x,x')
+{\rm O}(\varepsilon^{2}).
\label{(2.4)}
\end{equation}
Its existence is proved by the calculations described hereafter. Indeed,
by insertion of (2.4) into (2.3) we therefore obtain, picking out
terms of zeroth and first order in $\varepsilon$, the pair of
differential equations
\begin{equation}
\cstok{\ }^{0}G^{(0)}(x,x')=J^{(0)}(x,x'),
\label{(2.5)}
\end{equation}
\begin{equation}
\cstok{\ }^{0}G^{(1)}(x,x')=J^{(1)}(x,x'),
\label{(2.6)}
\end{equation}
having set
\begin{equation}
J^{(0)}(x,x') \equiv -\delta(x,x'),
\label{(2.7)}
\end{equation}
\begin{equation}
J^{(1)}(x,x') \equiv {z\over 2a}\delta(x,x')
-\left({z\over a}{\partial^{2}\over \partial t^{2}}
+{1\over 2a}{\partial \over \partial z} \right)G^{(0)}(x,x').
\label{(2.8)}
\end{equation}
Our boundary conditions are Dirichlet in the spatial variable $z$.
Since the full Feynman function $G_{F}(x,x')$ is required to vanish
at $z=0,a$, this implies the following homogeneous Dirichlet conditions
on the zeroth and first-order terms:
\begin{equation}
G^{(0)}(x,x') \biggr |_{z=0,a}=0,
\label{(2.9)}
\end{equation}
\begin{equation}
G^{(1)}(x,x') \biggr |_{z=0,a}=0.
\label{(2.10)}
\end{equation}
To solve Eqs. (2.5) and (2.6), we perform a Fourier analysis of 
$G^{(0)}$ and $G^{(1)}$, which remains meaningful in a weak gravitational
field \cite{PHRVA-D74-085011}, by virtue of translation invariance. 
In such an analysis we separate the
$z$ variable, i.e. we write (cf. \cite{PHRVA-D74-085011})
\begin{equation}
G^{(0)}(x,x')=\int {dk^{0}d{\vec k}_{\perp} \over (2\pi)^{3}}
\gamma^{(0)}(z,z')e^{i {\vec k}_{\perp} \cdot 
({\vec x}_{\perp}-{\vec x}_{\perp}')
-ik^{0}(x_{0}-x_{0}')},
\label{(2.11)}
\end{equation}
and similarly for $G^{(1)}(x,x')$, with a ``reduced Green function''
$\gamma^{(1)}(z,z')$ in the integrand as a counterpart of the zeroth-order
Green function $\gamma^{(0)}(z,z')$ in (2.9). Equations (2.3) and (2.4)
lead therefore to the following equations for reduced Green functions
(hereafter $\lambda \equiv \sqrt{k_{0}^{2}-k_{\perp}^{2}}$):
\begin{equation}
\left({\partial^{2}\over \partial z^{2}}+\lambda^{2} \right)
\gamma^{(0)}(z,z')=-\delta(z,z'),
\label{(2.12)}
\end{equation}
\begin{equation}
\left({\partial^{2}\over \partial z^{2}}+\lambda^{2} \right)
\gamma^{(1)}(z,z')
={z\over 2a}\delta(z,z')+\left({z\over a}k_{0}^{2}
-{1\over 2a}{\partial \over \partial z}\right)\gamma^{(0)}(z,z').
\label{(2.13)}
\end{equation}
By virtue of the Dirichlet conditions (2.9), $\gamma^{(0)}$ reads as
\begin{equation}
\gamma^{(0)}(z,z')=-{\sin(\lambda z_{<})\sin(\lambda(z_{>}-a))\over
\lambda \sin (\lambda a)},
\label{(2.14)}
\end{equation}
where $z_{<} \equiv {\rm min}(z,z')$,
$z_{>} \equiv {\rm max}(z,z')$. The evaluation of the reduced Green
function $\gamma^{(1)}$ is slightly more involved. For this purpose, we
distinguish the cases $z<z'$ and $z>z'$, and find the two equations
\begin{equation}
\left({\partial^{2}\over \partial z^{2}}+\lambda^{2} \right)
\gamma_{\pm}^{(1)}(z,z')=j_{\pm}^{(1)}(z,z'),
\label{(2.15)}
\end{equation}
where
\begin{equation}
j_{-}^{(1)}={1\over 2a}
{\lambda \cos (\lambda z)-2z k_{0}^{2}\sin (\lambda z) \over
\lambda \sin (\lambda a)} \sin (\lambda(z'-a)) \;
{\rm if} \; z< z',
\label{(2.16)}
\end{equation}
\begin{equation}
j_{+}^{(1)}={1\over 2a}{\lambda \cos (\lambda (z-a))
-2z k_{0}^{2} \sin (\lambda (z-a)) \over \lambda \sin (\lambda a)}
\sin (\lambda z') \; {\rm if} \; z > z'.
\label{(2.17)}
\end{equation}
We have therefore two different solutions in the intervals
$z < z'$ and $z > z'$. In this case the differential equation (2.15)
is solved by imposing the matching condition
\begin{equation}
\gamma_{-}^{(1)}(z',z')=\gamma_{+}^{(1)}(z',z')
\label{(2.18)}
\end{equation}
jointly with the jump condition
\begin{equation}
{\partial \over \partial z}\gamma_{+}^{(1)} \biggr |_{z=z'}
-{\partial \over \partial z}\gamma_{-}^{(1)} \biggr |_{z=z'}
={z' \over 2a}.
\label{(2.19)}
\end{equation}
Equation (2.18) is just the continuity requirement of the reduced
Green function $\gamma^{(1)}(z,z')$ at $z=z'$, 
while Eq. (2.19) can be obtained
by integrating Eq. (2.13) in a neighborhood of $z'$, since
\begin{equation}
\lim_{\epsilon \to 0}{\partial \over \partial z}\gamma^{(1)}
\biggr |_{z'-\epsilon}^{z'+\epsilon}
=\lim_{\epsilon \to 0}
\int_{z'-\epsilon}^{z'+\epsilon}{z\over 2a}\delta(z,z')dz
={z'\over 2a}.
\label{(2.20)}
\end{equation}
Bearing in mind Eq. (2.14) we can therefore write, for all $z,z'$,
\begin{eqnarray}
\gamma^{(1)}(z,z')&=& {1\over 4a \lambda^{2}} \biggr \{
\left[(k_{0}^{2}-\lambda^{2})(z+z')
-k_{0}^{2} \left(z^{2}{\partial \over \partial z}
+{z'}^{2}{\partial \over \partial z'}
\right)\right] \gamma^{(0)}(z,z')\nonumber \\
&-& k_{0}^{2}a^{2}{\sin (\lambda z) \sin (\lambda z') \over
\sin^{2}(\lambda a)} \biggr \}.
\label{(2.21)}
\end{eqnarray}

\section{Regularized and renormalized energy-momentum tensor}

In the previous section we have focused on the Feynman Green function
$G_{F}$ because it is then possible to develop a recursive scheme for
the evaluation of its asymptotic expansion at small $\varepsilon$.
However, we eventually need the Hadamard
function $H(x,x')$, which is obtained as \cite{PHRVA-D74-085011}
\begin{equation}
H(x,x') \equiv 2 {\rm Im}G_{F}(x,x') \sim 
2 {\rm Im}(G^{(0)}(x,x')+\varepsilon G^{(1)}(x,x'))
+{\rm O}(\varepsilon^{2}).
\label{(3.1)}
\end{equation}
The coincidence limits in the formula of the regularized and renormalized 
energy-momentum tensor 
make it necessary to perform the replacements
\begin{equation}
H_{;\mu' \nu}+H_{;\mu \nu'} \rightarrow P_{\mu}^{\; \mu'}H_{;\mu' \nu}
+P_{\nu}^{\; \nu'}H_{;\mu \nu'}, \;
H_{;\sigma}^{\; \; \; \sigma'} \rightarrow g^{\sigma \rho} 
P_{\rho}^{\; \rho'} H_{;\sigma \rho'}, \;
H_{;\mu' \nu'} \rightarrow P_{\mu}^{\; \mu'} P_{\nu}^{\; \nu'}
H_{;\mu' \nu'},
\label{(3.2)}
\end{equation}
where $P_{\; \nu'}^{\mu}$ is the parallel displacement 
bivector \cite{PHRVA-D77-105011}
\begin{equation}
P_{\; \nu'}^{\mu} \sim {\rm diag}
\left(1+{\varepsilon \over 2a}(z'-z),1,1,1 \right)
+{\rm O}(\varepsilon^{2}).
\label{(3.3)}
\end{equation}
Hence we get the asymptotic expansion at small $\varepsilon$ of the
regularized energy-momentum tensor according to (hereafter we
evaluate its covariant, rather than contravariant, form)
\begin{equation}
\langle T_{\mu \nu} \rangle \sim \langle T_{\mu \nu}^{(0)} \rangle
+\varepsilon \langle T_{\mu \nu}^{(1)} \rangle 
+{\rm O}(\varepsilon^{2}),
\label{(3.4)}
\end{equation}
where, on defining $s \equiv \pi z /a, \; s' \equiv \pi z'/a$, we find
\begin{eqnarray}
\langle T_{\mu \nu}^{(0)} \rangle &=& \left[
-{\pi^{2}\over 1440 a^{4}}-\lim_{s' \to s}
{\pi^{2}\over 2a^{4}(s-s')^{4}}\right]
\begin{pmatrix}
1 & 0 & 0 & 0 \\
0 & -1 & 0 & 0 \\
0 & 0 & -1 & 0 \\
0 & 0 & 0 & 3
\end{pmatrix}
\nonumber \\
&+& \left(\xi -{1\over 6} \right){\pi^{2} \over 8a^4}
\left[{3 -2 \sin^{2}s \over \sin^{4}s}\right]
\begin{pmatrix}
1 & 0 & 0 & 0 \\
0 & -1 & 0 & 0 \\
0 & 0 & -1 & 0 \\
0 & 0 & 0 & 0
\end{pmatrix},
\label{(3.5)}
\end{eqnarray}
and
\begin{eqnarray}
\langle T_{00}^{(1)} \rangle &=& 
{\pi \over 1440 a^{4} \sin^{4}s} \biggr[{311 \over 40}\pi
-{637 \over 40}s +{1\over 10}(43 \pi -81s)\cos 2s \nonumber \\
&+& {s -3\pi \over 40}\cos 4s +5 \sin 2s +2(\pi -s)s 
(\sin 2s -6 \cot s)\biggr] \nonumber \\
&+& \left(\xi -{1\over 6}\right){\pi \over 48 a^{4}\sin^{4}s}
\Bigr[2(\pi+s)(2+\cos 2s)+{5\over 2} \sin 2s \nonumber \\
&+& (\pi-s)s (\sin 2s -6 \cot s)\Bigr]
-\lim_{s' \to s}{\pi s \over 2a^{4}(s-s')^{4}},
\label{(3.6)}
\end{eqnarray}
\begin{eqnarray}
\langle T_{11}^{(1)} \rangle &=& {\pi \over 7200 a^{4}}
\biggr[\pi-2s +{5\over \sin^{2}s} \Bigr(2(\pi-2s)
\left(-2+{3\over \sin^{2}s}\right) \nonumber \\
&+& \cot s \left(5+2(\pi -s)s -6(\pi -s){s\over \sin^{2}s}\right)
\Bigr) \biggr] \nonumber \\
&+& \left( \xi-{1\over 6} \right) {\pi \over 96a^{4} \sin^{5}s}
\Bigr[(11(\pi-s)s-1)\cos s \nonumber \\
&+& ((\pi-s)s+1)\cos 3s-2(\pi-2s)(3 \sin s +\sin 3s)\Bigr],
\label{(3.7)}
\end{eqnarray}
\begin{equation}
\langle T_{22}^{(1)} \rangle = \langle T_{11}^{(1)} \rangle ,
\label{(3.8)}
\end{equation}
\begin{equation}
\langle T_{33}^{(1)} \rangle=-{\pi^{2}\over 1440 a^{4}}
+{\pi s \over 720 a^{4}} +\left(\xi -{1\over 6}\right)
{\pi \over 16 a^{4}}{\cos s \over \sin^{3} s}.
\label{(3.9)}
\end{equation}

The next step of our analysis is the renormalization of the regularized
energy-momentum tensor. For this purpose, following our work in
Ref. \cite{PHRVA-D74-085011}, we subtract the energy-momentum tensor 
evaluated in the absence of bounding plates, i.e.
\begin{equation}
\langle {\widetilde T}_{\mu \nu}^{(0)} \rangle 
=-\lim_{s' \to s}{\pi^{2}\over 2a^{4}(s-s')^{4}}
\begin{pmatrix}
1 & 0 & 0 & 0 \\
0 & -1 & 0 & 0 \\
0 & 0 & -1 & 0 \\
0 & 0 & 0 & 3
\end{pmatrix},
\label{(3.10)}
\end{equation}
and
\begin{equation}
\langle {\widetilde T}_{\mu \nu}^{(1)} \rangle 
=-\lim_{s' \to s}{\pi s \over 2a^{4}(s-s')^{4}}
\begin{pmatrix}
1 & 0 & 0 & 0 \\
0 & 0 & 0 & 0 \\
0 & 0 & 0 & 0 \\
0 & 0 & 0 & 0
\end{pmatrix}.
\label{(3.11)}
\end{equation}
To test consistency of our results we should now check whether our
regularized and renormalized energy-momentum tensor is covariantly 
conserved, since otherwise we would be outside the realm of quantum
field theory in curved spacetime, which would be unacceptable. Indeed,
the condition
\begin{equation}
\nabla^{\mu} \langle T_{\mu \nu} \rangle =0
\label{(3.12)}
\end{equation}
yields, working up to first order in $\varepsilon$, the pair of equations
\begin{equation}
{\partial \over \partial z} \langle T_{33}^{(0)} \rangle=0,
\; (\varepsilon^{0} \; {\rm term})
\label{(3.13)}
\end{equation}
\begin{equation}
{\partial \over \partial z} \langle T_{33}^{(1)} \rangle
+{1\over 2a}\Bigr(\langle T_{00}^{(0)} \rangle 
+ \langle T_{33}^{(0)} \rangle \Bigr)=0 \;
(\varepsilon^{1} \; {\rm term}),
\label{(3.14)}
\end{equation}
which are found to hold identically for all values of $\xi$ 
in our problem.

The trace of $\langle T_{\mu \nu} \rangle$ is obtained as
\begin{equation}
\tau \equiv g^{\mu \nu} \langle T_{\mu \nu} \rangle
\sim \eta^{\mu \nu} \langle T_{\mu \nu}^{(0)} \rangle
+\varepsilon \Bigr[\eta^{\mu \nu} \langle T_{\mu \nu}^{(1)}
\rangle +{z\over a} \langle T_{00}^{(0)} \rangle \Bigr]
+{\rm O}(\varepsilon^{2}),
\label{(3.15)}
\end{equation}
from which we find a $\xi$-dependent part
\begin{eqnarray}
\tau_{\xi}&=& \left(\xi-{1\over 6}\right)\biggr \{ 
-{3\pi^{2}(2+\cos 2s)\over 8a^{4}\sin^{4}s}
-\varepsilon {\pi \over 32 a^{4}\sin^{5}s}
\Bigr[(1-11(\pi -s)s)\cos s \nonumber \\
&-& (1+(\pi-s)s)\cos 3s +2(\pi-2s)(3 \sin s + \sin 3s)\Bigr]
\biggr \}.
\label{(3.16)}
\end{eqnarray}
Interestingly, the value $\xi={1\over 6}$ which yields conformal
invariance of the classical action 
is the same as the value of $\xi$ yielding no trace anomaly
\cite{PHRVA-D77-105011}.

\section{Casimir energy and pressure}

In order to evaluate the energy density $\rho$ of our ``scalar'' Casimir 
apparatus, we project the regularized and renormalized
energy-momentum tensor along a unit timelike vector
$u^{\mu}=\left(-{1\over \sqrt{-g_{00}}},0,0,0 \right)$. This yields
\begin{eqnarray}
\rho &=& \langle T_{\mu \nu} \rangle u^{\mu} u^{\nu}
=-{\pi^{2}\over 1440 a^{4}}+{\pi \over 7200 a^{4}}
\biggr[-3 \pi+6s +{10\over \sin^{2}s}\Bigr(2(\pi-2s) \nonumber \\
& \times & \left(-2+{3\over \sin^{2}s}\right)
+\cot s \left((5+2(\pi-s)s+6{s(-\pi+s)\over 
\sin^{2}s}\right)\Bigr)\biggr]
\varepsilon \nonumber \\
&+& \left(\xi-{1\over 6}\right) \biggr \{ 
{\pi^{2}(2+\cos 2s)\over 8a^{4} \sin^{4}s}
-{\pi \over 192 a^{4} \sin^{5}s}\biggr[\Bigr(-5+22(\pi-s)s\Bigr)
\cos s \nonumber \\
&+& \Bigr(5+2(\pi-s)s \Bigr)\cos 3s -4(\pi-2s)(3 \sin s +\sin 3s)
\biggr] \varepsilon \biggr \}.
\label{(4.1)}
\end{eqnarray}
The energy $E$ stored within our Casimir cavity is given by
\begin{equation}
E=\int_{V_{c}}d^{3}\Sigma \sqrt{-g}\rho,
\label{(4.2)}
\end{equation}
where $d^{3}\Sigma$ is the volume element of an observer with
four-velocity $u^{\mu}$, and $V_{c}$ is the volume of the cavity.
The integration used here requires the use of approximating domains, 
i.e. the $z$-integration is performed in the interval
$(\zeta,a-\zeta)$, corresponding to ${\pi \over a}(\zeta,a-\zeta)$ 
in the $s$ variable, taking eventually
the $\zeta \to 0$ limit. We thus obtain \cite{PHRVA-D77-105011}
\begin{equation}
E_{\xi}=-{\pi^{2}A\over 1440 a^{3}}
-{\pi^{2}A \varepsilon \over 5760 a^{3}}
+\left(\xi -{1\over 6}\right){\pi A \over 4a^{3}}
\left(1+{\varepsilon \over 4}\right)
\lim_{\zeta \to 0}{\cos \zeta \over \sin^{3}\zeta},
\label{(4.3)}
\end{equation}
where $A$ is the area of parallel plates. Note that the conformal coupling
value $\xi={1\over 6}$ is picked out as the only value of $\xi$ for which
the Casimir energy remains finite. In this case, reintroducing the constants
$\hbar,c$ and writing explicitly $\varepsilon$, 
we find \cite{PHRVA-D77-105011}
\begin{equation}
E_{c}=-{\hbar c \pi^{2}\over 1440}{A \over a^{3}}
\left(1+{1\over 2}{ga \over c^{2}}\right).
\label{(4.4)}
\end{equation}

In the same way, the pressure $P_{\xi}$ on the parallel plates is
found to be \cite{PHRVA-D77-105011}
\begin{equation}
P_{\xi}(z=0)={\pi^{2}\over 480 a^{4}}
+{\pi^{2}\varepsilon \over 1440 a^{4}}
-\left(\xi-{1\over 6}\right){\pi \varepsilon \over 16a^{4}}
\lim_{s \to 0}{\cos s \over \sin^{3}s},
\label{(4.5)}
\end{equation}
\begin{equation}
P_{\xi}(z=a)=-{\pi^{2}\over 480 a^{4}}
+{\pi^{2}\varepsilon \over 1440 a^{4}}
+\left(\xi-{1\over 6}\right){\pi \varepsilon \over 16a^{4}}
\lim_{s \to \pi}{\cos s \over \sin^{3}s}.
\label{(4.6)}
\end{equation}
Once again, one can get rid of divergent terms by setting 
$\xi={1\over 6}$, which leads to \cite{PHRVA-D77-105011}
\begin{equation}
P_{c}(z=0)={\pi^{2}\over 480} {\hbar c \over a^{4}}
\left(1+{2\over 3}{ga \over c^{2}}\right),
\label{(4.7)}
\end{equation}
\begin{equation}
P_{c}(z=a)=-{\pi^{2}\over 480}{\hbar c \over a^{4}}
\left(1-{2\over 3}{ga \over c^{2}}\right).
\label{(4.8)}
\end{equation}
To obtain the resulting force one has to multiply each of these
pressures by the redshift $r$ of the point where they act, relative
to the point where they are added \cite{PHRVA-D76-025008}, i.e.,
\begin{equation}
r=\sqrt{|g_{00}(P_{\rm act})| \over 
|g_{00}(P_{\rm added})|} \approx 1+{g\over c^{2}}(z-z_{Q}),
\label{(4.9)}
\end{equation}
to leading order in ${gz \over c^{2}}$. Thus, a net force 
is obtained of magnitude
\begin{equation}
F=-{\pi^{2} \hbar c \over a^{4}}\left[{g\over 480 c^{2}}(z_{2}-z_{1})
-{4g \over 1440 c^{2}}(z_{2}-z_{1})\right]
={\pi^{2}\over 1440}{A \hbar g \over c a^{3}}
={| E_{C}^{0}| \over c^{2}}g,
\label{(4.10)}
\end{equation}
having defined $E_{C}^{0} \equiv -{\pi^{2}\over 1440}\hbar c 
{A\over a^{3}}$, which points upwards along the $z$-axis and is in
full agreement with the equivalence principle.  

\section{Neumann boundary conditions}

When the reduced Green functions obey instead Neumann boundary conditions
on parallel plates, i.e.
\begin{equation}
{\partial \gamma^{(i)}\over \partial z}\biggr |_{z=0}
={\partial \gamma^{(i)}\over \partial z}\biggr |_{z=a}=0, \;
\forall i=0,1, \;
\label{(5.1)}
\end{equation}
our work in Ref. \cite{PHRVA-D78-107701} has found, by an analogous 
procedure, the regularized and renormalized energy-momentum tensor
to first order in $\varepsilon$, with trace
\begin{eqnarray}
\; & \; & \tau_{\xi} \equiv g^{\mu \nu} \langle T_{\mu \nu} \rangle
\nonumber \\
&=& \left(\xi -{1\over 6}\right){\pi \over 32a^{4}}
{1\over \sin^{5}s} \Bigr \{ 6\pi (3 \sin s + \sin 3s) 
\nonumber \\
&-& \varepsilon \Bigr[(1+11(\pi-s)s)\cos s -(1-(\pi -s)s)
\cos 3s \nonumber \\ 
&-& 2(\pi-2s)(3 \sin s +\sin 3s)\Bigr] \Bigr \},
\label{(5.2)}
\end{eqnarray}
which vanishes in the case of conformal coupling, as with Dirichlet
boundary conditions. Moreover, the Casimir energy stored between the
plates, the pressure on parallel plates and the net force are then
found to agree completely with (4.4), (4.5)--(4.6) and (4.10),
respectively.

\section{Electromagnetic field}

The work in Ref. \cite{PHRVA-D78-024010} has instead exploited the
fact that, to first order in the small quantity $gz$, the line
element (1.1) coincides with the Rindler metric
\begin{equation}
ds^{2}=-\left({\xi \over \xi_{1}}\right)^{2}dt^{2}+d\xi^{2}
+dx_{\perp}^{2},
\label{(6.1)}
\end{equation}
where $\xi \equiv {1\over g}+z \equiv \xi_{1}+z$. In a fully covariant
analysis of Feynman Green functions in Rindler spacetime, the 
components of the Maxwell energy-momentum tensor have been evaluated
up to second order in $g$ with perfect conductor boundary conditions, 
finding that, as $z \rightarrow 0$,
\begin{equation}
\langle 0 | T_{t}^{\; t}| 0 \rangle \sim
{g \over 30 \pi^{2} z^{3}}+{\rm O}(z^{-2}),
\label{(6.2)}
\end{equation}
\begin{equation}
\langle 0 | T_{z}^{\; z}| 0 \rangle \sim
-{g^{2}\over 60 \pi^{2}z^{2}}+{\rm O}(z^{-1}),
\label{(6.3)}
\end{equation}
\begin{equation}
\langle 0 | T_{x}^{\; x}| 0 \rangle 
=\langle | T_{y}^{\; y}| 0 \rangle
\sim -{g \over 60 \pi^{2}z^{3}}+{\rm O}(z^{-2}).
\label{(6.4)}
\end{equation}
Since $T_{zz}$ is now found to diverge on approaching the plates, no 
definite meaning can be given to the gravitational correction to the
Casimir pressure. Moreover, the divergences in $T_{t}^{\; t}$ are such
that the resulting correction to the total mass energy of the 
cavity is infinite, even on taking the principal-value integral 
of $T_{t}^{\; t}$. 

\section{Concluding remarks}

The literature on the behaviour of rigid Casimir cavities in a weak
gravitational field predicts, on theoretical ground, that Casimir 
energy obeys exactly the equivalence principle, and hence the Casimir
apparatus should experience a tiny push in the upward direction. The
formula for the push has been obtained in three different ways, i.e.
a heuristic summation over modes \cite{PHLTA-A297-328, IMPAE-A17-804},
or a variational approach \cite{PHRVA-D76-025004, JPAGB-40-10935},
or an energy-momentum analysis \cite{PHRVA-D74-085011, PHRVA-D78-024010}.
Moreover, the work in Ref. \cite{08100081} has shown that Casimir energy
for a configuration of parallel plates gravitates according to the 
equivalence principle both for the finite and divergent parts. This
suggests that such divergent parts can be absorbed by a process of
renormalization \cite{08100081}.

It now remains to be seen whether this interpretation is viable in all
configurations of physical interest. Moreover, on the experimental side,
the signal-modulation problems first discussed in Refs. 
\cite{PHLTA-A297-328, IMPAE-A17-804} remain, to our knowledge, unsolved,
while being of extreme importance on studying the feasibility of the
experiment.

Last, but not least, our findings should be compared with those in
Ref. \cite{08012063}, where the authors consider the cosmological
evolution in a recently suggested new model of quantum initial 
conditions for the Universe. They find that the effective Friedmann
equation incorporates the effect of the conformal anomaly of quantum 
fields, and shows that their vacuum Casimir energy is completely 
screened and {\it does not gravitate}.

\acknowledgments
G. Esposito is grateful to the Dipartimento di Scienze Fisiche of 
Federico II University, Naples, for hospitality and support.
The authors are grateful to Steve Fulling and Kim Milton for 
correspondence, and to Aram Saharian for scientific conversations.

\end{document}